\documentclass[english]{iopart}
\usepackage[T1]{fontenc}
\usepackage[latin9]{inputenc}
\usepackage{textcomp}
\usepackage{graphicx}
\usepackage{todonotes}

\usepackage[unicode=true,pdfusetitle,
 bookmarks=true,bookmarksnumbered=false,bookmarksopen=false,
 breaklinks=false,backref=false,colorlinks,
 citecolor=blue, linkcolor=blue]
 {hyperref}

\makeatletter

\usepackage{iopams}
\usepackage{setstack}

\makeatother

\usepackage{babel}
\begin{document}
\title{Formation of the frozen core in critical Boolean Networks}

\author{Marco M\"oller and Barbara Drossel}

\address{Institute for condensed matter physics, TU Darmstadt, Hochschulstrasse 6, 64289 Darmstadt, Germany }

\begin{abstract}
We investigate numerically and analytically the formation of the
frozen core in critical random Boolean networks with biased
functions. We demonstrate that a previously used efficient algorithm
for obtaining the frozen core, which starts from the nodes with
constant functions, fails when the number of inputs per node exceeds
4. We present computer simulation data for the process of formation of
the frozen core and its robustness, and we show that several important
features of the data can be derived by using a mean-field calculation.
\end{abstract}
\maketitle

\section{Introduction}
Boolean networks are often used as generic models for the dynamics of
complex systems of interacting entities, such as social and economic
networks, neural networks, and gene or protein interaction networks
\cite{kauffman:random,kauffman:metabolic}. Whenever the states of a
system can be reduced to being either ``on'' or ``off'' without loss
of important information, a Boolean appriximation captures many
features of the dynamics of real networks \cite{BornholdtScience}. In
order to understand the generic behavior of such models, random models
were investigated in depth \cite{reviewbarbara}, although it is clear
that neither the connection pattern, nor the usage of update functions
of biological networks is reflected realistically in such random
models.  Most recent research on Boolean networks has therefore been
devoted to networks with more realistic features, however, there
remain important open questions concerning the behavior of random
models.

In a Random Boolean model, the connections and the update functions
are assigned to the nodes at random, given the number of nodes $N$, the
number of inputs per node $k$, and the probability distribution for the
update functions. Dynamics are usually implemented by updating the
nodes of the network in parallel. Starting from some initial
configuration, the system eventually settles on a periodic attractor.
Of special interest are \emph{critical} networks, which lie at the
boundary between a frozen phase and a chaotic phase
\cite{derrida:random,derrida:phase}.  In the frozen phase, a
perturbation at one node propagates during one time step on an average
to less than one node, and the attractor lengths remain finite in the
limit of infinite node number $N\to \infty$. In the
chaotic phase, the difference between two almost identical states
increases exponentially fast, because a perturbation propagates on an
average to more than one node during one time step
\cite{aldana-gonzalez:boolean}. Whether a network is frozen, chaotic,
or critical, depends on the in-degree $k$ as well as on the weights of
the different Boolean functions. If these weights are chosen
appropriately, critical networks can be created for any value of $k$.
 
In order to gain a deeper understanding of the critical behavior, it
has proven useful to classify the nodes according to their behavior on
an attractor. First, there are nodes that are frozen on the same value
on every attractor. Such nodes give a constant input to other nodes
and are otherwise irrelevant. They form the \emph{frozen core} of the
network \cite{flyvbjerg:order}. Second, there are nodes whose outputs
go only to irrelevant nodes. Though they may fluctuate, they are also
classified as irrelevant since they act only as slaves to the nodes
determining the attractor period. Third, the \emph{relevant nodes} are
the nodes whose state is not constant on all attractors and that
control at least one relevant node. These nodes determine completely
the number and period of attractors.  The recognition of the relevant
elements as the only elements influencing the asymptotic dynamics was
an important step in understanding the attractors of Kauffman networks
\cite{flyvbjerg:exact,bastolla:relevant,bastolla:modular,socolar:scaling,kaufmanandco:scaling,TamaraContainerAnalytisch}.
Due to these publications, it is now established that in critical
Random Boolean networks the mean number of nonfrozen nodes scales as
$N_{nf}\sim N^{2/3}$, and the mean number of relevant nodes scales as
$N_{rel}\sim N^{1/3}$ when $k>1$. The frozen core thus comprises all
but of the order of $N^{2/3}$ nodes.

 In order to determine the frozen core, an efficient algorithm has
 been suggested in
 \cite{kaufmanandco:scaling,TamaraContainerAnalytisch}, which starts
 from the nodes with constant functions and determines iteratively all
 other nodes which become frozen due to being influenced by other
 frozen nodes. The advantage of this approach is that one can explore
 numerically very large networks, which would not be accessible to a
 direct modelling approach. Furthermore, this algorithm could be
 translated into a stochastic process. From the Fokker-Planck equation
 of this process it was possible to derive the above-mentioned scaling
 laws and many other results analytically. A third advantage of this
 approach is that it explains naturally why all but of the order of
 $N^{2/3}$ nodes become frozen on the same value for all initial conditions.

There exist also other mechanisms that cause the freezing of nodes.
Small network motifs where two or more paths lead from the same
``initial'' to the same ``final'' node can freeze the final node when
the update functions of all nodes in the motif are chosen
appropriately. However, such motifs occur only in very small numbers
in random networks and do not play an important role. Of greater
importance are loops of nodes with canalyzing functions that can fix
each other on their canalyzing value. Such loops are called ``forcing
loops'' in \cite{kauffman1984emergent} and ``self-freezing
loops'' in \cite{paul2006properties}. In that paper, it was shown that a
canalyzing random Boolean network with $k=2$, which has no
constant functions, nevertheless has a frozen core due to these
self-freezing loops. This is, however, a special mechanism that occurs
only for very specific sets of Boolean functions. 

In this paper, we will first show that the assumption that the frozen
core can be obtained by starting from the nodes with constant
functions becomes wrong for sufficiently large values of $k$. When
biased update functions are used, the method outlined in the previous
paragraph fails for values of $k$ larger than 4. When other sets of
update functions are used, the method can already fail for $k=
3$. Therefore, we will study the process of the formation of the frozen
core in more depth, and we will show that for larger $k$ the frozen
core has features very similar to those of systems with smaller $k$,
despite the fact that the frozen core cannot be obtained any more by
starting from constant functions. We will perform computer simulations
as well as present analytical considerations in order to corroborate
our findings.

\section{Model}

A Random Boolean network consists of $N$ nodes, each of which receives
input from $k$ randomly chosen other nodes. Furthermore, each node is
assigned a Boolean update function.  We choose \emph{biased functions},
which are characterized by a parameter $p$, assigning to
each of the $2^{k}$ input configurations the output 1 with a
probability $p$ and the output 0 with a probability $1-p$. The value
of $p$ is chosen such that the network is
critical \cite{obsolete_aldana-gonzalez:boolean},
\begin{equation}
p=\frac{1}{2}\pm\sqrt{\frac{1}{4}-\frac{1}{2k }}.\label{eq:Pcrict}
\end{equation}
In the following, we will use only the minus sign, which means that
our update functions are dominated by the value $0$ and that $p$
is the probability of the minority bit, which is 1.

Occasionally, we will also refer to other sets of functions, but all our computer simulations were done with biased functions.

\section{Determining the frozen core starting from constant functions}

An elegant way to determine the frozen core was suggested in
\cite{kaufmanandco:scaling,TamaraContainerAnalytisch}.  This method is
based on the assumption that allmost all frozen nodes can be obtained
by starting from the nodes with a constant update function and by
determining iteratively all nodes that become frozen because some of
their inputs are frozen. The algorithm can be formulated as a
stochastic process which places all nodes in containers $C_{i}$
without yet specifying their functions or their connections. We denote
the number of nodes in container $C_{i}$ by $N_i$. The index $i$
denotes the number of nonfrozen inputs of the nodes in container
$C_{i}$. Initially, each node is placed in container $C_{0}$ with a
probability
\begin{equation}
\beta\left(k\right) \equiv p^{2^k}+(1-p)^{2^k}
\end{equation}
 (which is the probability that it has a
constant function), and in container $C_{k}$ otherwise.  The algorithm
then proceeds by selecting one node from container $C_{0}$ and
evaluating to which other nodes it is an input. This evaluation is
done by connecting the $i$ nonfrozen inputs of each node in the
containers with $i>0$ with a probability $1/\sum_i N_i$ to the
selected node. All nodes for which $m>0$ inputs become connected to
the selected node, are then moved from their original container
$C_{i}$ to container $C_{i-m}$ -- unless they become completely frozen
because the outputs are a constant function of the remaining nonfrozen
inputs. In this case, a node is moved to container $C_{0}$ instead of
$C_{i-m}$. The probability that a node in container $C_{i}$ becomes frozen when one of its inputs is frozen, is
\begin{equation}
\omega_{i}  = \frac{\beta\left(i-1\right)-\beta\left(i\right)}{1-\beta\left(i\right)} \, .\label{omega}
\end{equation}
When $m>1$, one evaluates for one input after the other whether its fixation
causes the node to become frozen. 

After determining all nodes to which the selected node is an input and
moving these nodes to the appropriate containers, the selected node is
removed from container $C_{0}$, and $N_0$ is reduced by 1. The total
number of nodes in the containers thus decreases by 1 during each
iteration.  The algorithm stops iff $N_0=0$. The remaining nodes are
those then supposed to be not part of the frozen core.

\begin{figure}
\includegraphics[width=1\columnwidth]{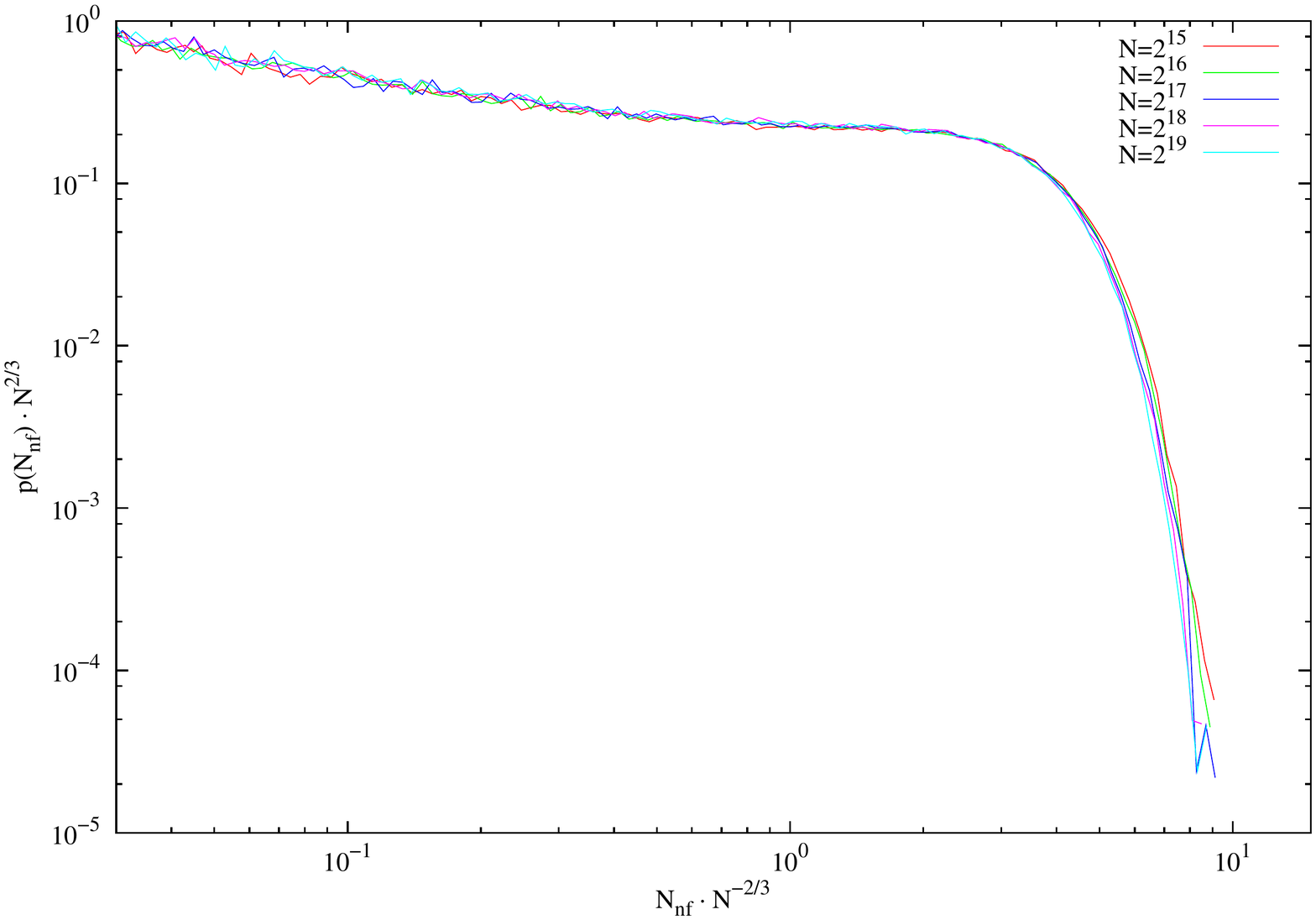}

\caption{\label{fig:ScalecollapsFrozCore} Probability distribution of
  the number $N_{nf}$ of non frozen nodes for $k=4$, for different
  $N\in\{2^{15},\ldots,2^{19}\}$, and scaled with $N^{2/3}$. The data
  were obtained by using the container method, which determines the
  frozen core by starting from the nodes with constant
  functions. 100\,000 randomly choosen networks where evaluated for
  each value of $N$.}
\end{figure}

Figure~\ref{fig:ScalecollapsFrozCore} shows the probability
distribution for the number of nonfrozen nodes obtained for $k=4$
with this method. The data for different network sizes $N$ are
collapsed to one universal curve by scaling with $N^{2/3}$. In
agreement with the general theory presented in
\cite{kaufmanandco:scaling,TamaraContainerAnalytisch}, the scaling
function is independent of $k$ and appears identical to the one
presented in \cite{kaufmanandco:scaling} for $k=2$.

However, when this procedure is performed for $k\ge 5$, it fails. Only a
small part of all nodes become frozen by starting from the nodes with
constant functions.

In order to understand this failure of the procedure for $k \ge 5$,
let us consider the deterministic difference equations that describe
the stochastic process of the container method as long as all $N_i$
are large. We denote with $N(t)$ the total number of nodes in the containers at step $t$. At the beginning we have 
\begin{eqnarray}
N\left(0\right) & = & N \label{container1}\\
N_{0}\left(0\right) & = & \beta\left(k\right) N\\
N_{i}\left(0\right) & = & 0\\
N_{k}\left(0\right) & = & \left(1-\beta\left(k\right)\right) N\, .
\end{eqnarray}
During each step,  the mean container contents change according to
\begin{eqnarray}
\Delta N\left(t\right) & = & -1\\
\Delta N_{0}\left(t\right) & = & -1+\sum_{i>0}\omega_{i}\left(i+1\right)\frac{N_{i+1}\left(t\right)}{N\left(t\right)}\\
\Delta N_{i}\left(t\right) & = & -i\frac{N_{i}\left(t\right)}{N\left(t\right)}+\left(1-\omega_{i}\right)\left(i+1\right)\frac{N_{i+1}\left(t\right)}{N\left(t\right)}\\
\Delta N_{k}\left(t\right) & = & -k\frac{N_{k}\left(t\right)}{N\left(t\right)},\label{container4}
\end{eqnarray}
as long as $N_0>0$. If $N_0$ reaches $0$, the process stops.  This is
valid for sufficiently large networks where the probability that two
nodes are connected by more than one edge vanishes. For
$N\rightarrow\infty$, one can replace these difference equations by
differential equations.

\begin{figure}
\includegraphics[width=1\columnwidth]{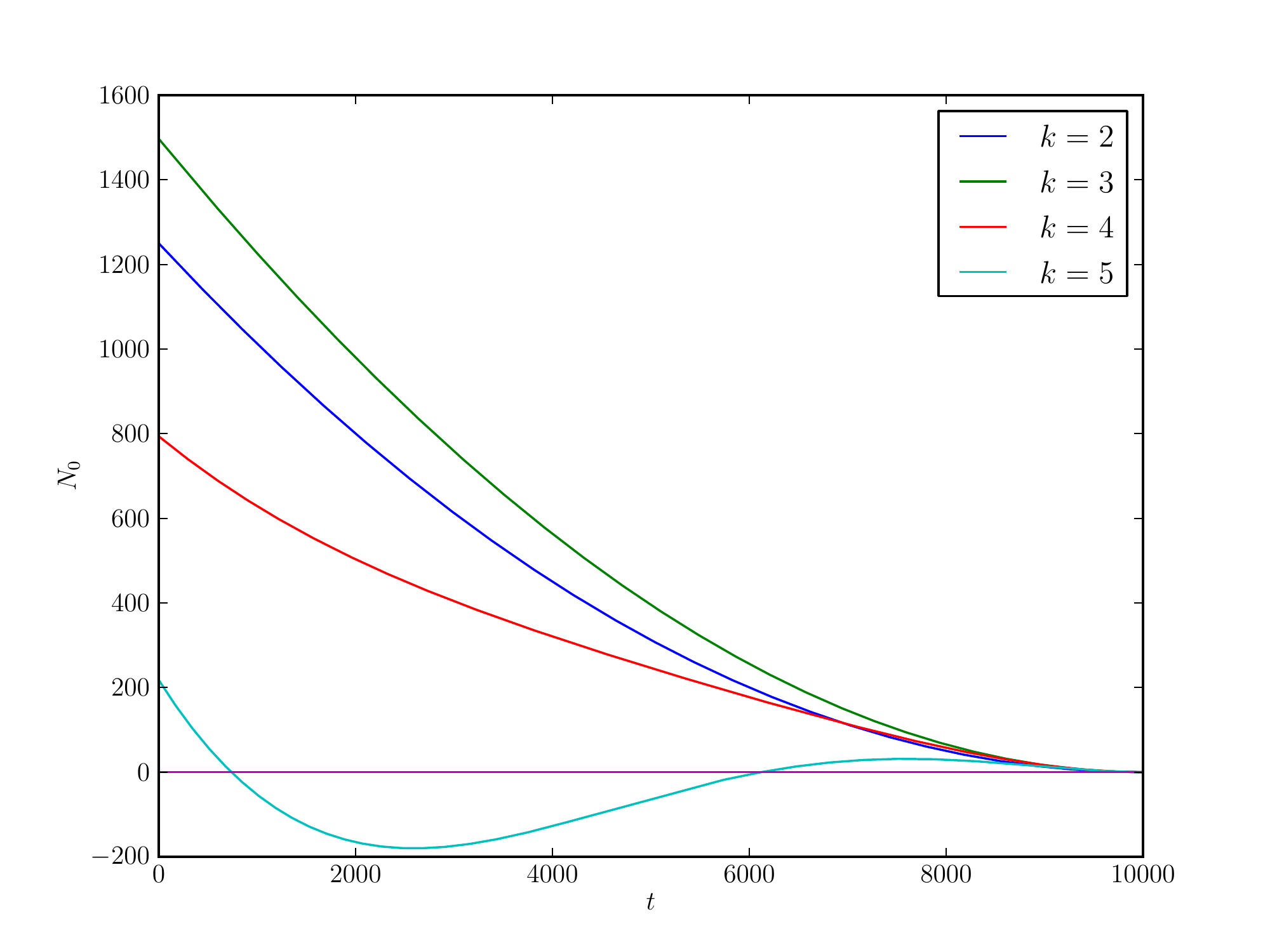}

\caption{\label{fig:ContainerDet} Number of nodes $N_0(t)$ in container $C_0$
obtained by iterating the deterministic equations (\ref{container1}-\ref{container4}) for different values of $k$ and network size $N=10\,000$. We continued the computation until $t=N$. For $k > 4$, the function $N_{0}(t)$ intersects the $x$-axis twice.}
\end{figure}

Figure~\ref{fig:ContainerDet} shows the number of nodes in container
$C_{0}$ for different values of $k$ for a numerical iteration of the
deterministic difference equations. For $k=5$, we did not stop the iteration at $N_0=0$, but we continued to $t=N$. For $k<5$, $N_0$ decreases monotonically
from its initial value at $t=0$ to 0 at $t=N$. For
$k = 5$, $N_0$ becomes negative for a value $t/N$ close to 0 and
becomes positive again only when $t/N$ is not too far from the value
1. From there, it reaches a local maximum and decreases then again to
0 at $t/N=1$. A similar behavior is found for larger $k$ (not shown).

The fact that $N_0$ becomes zero while a large proportion of the
network is not yet frozen means that the frozen core cannot be built
by starting only from those nodes that have constant functions. In
principle, this could also mean that different sets of nodes remain
unfrozen for different initial conditions, or that those nodes that
freeze for all initial conditions do not always freeze on the same
value. However, as we will argue below, the frozen core comprises also
for $k \ge 5$ all but of the order $N^{2/3}$ nodes of the network.  If
we want to interpret the fact that $N_0$ first becomes negative and
then becomes positive again at a larger $t/N$, we can reason as
follows: When we continue freezing inputs and decreasing $N_0$ by 1 at
each step, even though $N_0$ is 0 or negative, we assume that there
exist frozen nodes that we have not yet identified, but that we will
identify later. This assumption does not lead to a contradiction if
during this process enough nodes freeze that $N_0$ becomes again
positive. The assumption that there exists a large number of frozen
nodes that are not frozen by starting from the constant functions, is
then proven self-consistent.

When a more general set of update functions is used instead of biased
functions, the condition $N_0(t)=0$ can have nontrivial solutions
already for $k=3$. For a general set of functions
\cite{TamaraContainerAnalytisch}, the probabilities $\omega_i$ that a
node with $i +1 $ inputs becomes frozen when one of its inputs
freezes, can take values different from those for biased functions,
Equation~(\ref{omega}).  For $k=3$, we obtain the following conditions
for the existence of a solution $N_0(t)=0$ for $t/N < 1$:
\begin{eqnarray*}
0<\omega_{2} & < & \frac{1}{3}\\
\frac{1}{2}<\omega_{1} & < & 1-\frac{1}{3\left(1-\omega_{2}\right)}\\
1- \beta\left(3\right) & = & \frac{1}{3\left(1-\omega_{1}\right)\left(1-\omega_{2}\right)}.
\end{eqnarray*}
The third condition is the condition for criticality.

As an aside, we note that if $N_0$ becomes negative for critical
network, it will also become negative for ``frozen'' networks as long
as the control parameter is not too far away from its critical
value. This means that even networks that are in the frozen phase are
not necessarily frozen because of constant functions.

On the other hand, there exist sets of update functions where the
container method does not fail for any value of $k$.  
One such set is obtained by only choosing constant and reversible functions, as was done in \cite{drossel2009critical}. 
Another such set was used in \cite{samuelsson2006exhaustive}, where the process of formation of the
frozen core was viewed as an exhaustive bond percolation process.  By
assuming that the probability that a randomly chosen node is frozen if
$l$ of its inputs are frozen does not change during the process, the
authors obtained a simpler recursion relation than our difference
equations above. This simpler recursion relation is valid for bond
percolation, but not for our model with biased functions, where the set of
nodes that have already been removed from the system has a different
probability distribution of update functions than the set of nodes
that are still left in the containers. Therefore, we must
keep track of the probability distribution of the different types of
functions by monitoring the contents of each container.
 
Until now, we have focused on the size of the frozen core. For $k<5$
and biased functions, the frozen core can be determined by starting
from the constant nodes and determining iteratively all nodes that
become frozen because they have frozen inputs. However, the container
method does not reflect the real freezing dynamics. Therefore, we
studied the freezing process by computer simulations of networks. In
order to determine the influence of the nodes with constant functions
on the freezing dynamics, we compared the number of frozen nodes
obtained from a straightforward computer simulation of the network
with the number of nodes that become frozen because their inputs have
become frozen, which is the situation considered in the container
method.  In the first case, all nodes that did not change their state
during the remainder of the simulation were considered frozen after
the moment when they changed last. The freezing process according to
the ``container method'' was implemented by considering all nodes with
constant functions frozen at $t=0$, and by freezing at time $t$ all
those nodes that become frozen because one or more of their inputs
became frozen at time $t-1$. This amounts to running the container
method with a parallel update procedure, where all nodes in container
$C_0$ are dealt with during the same time step.

\begin{figure}
\includegraphics[width=1\columnwidth]{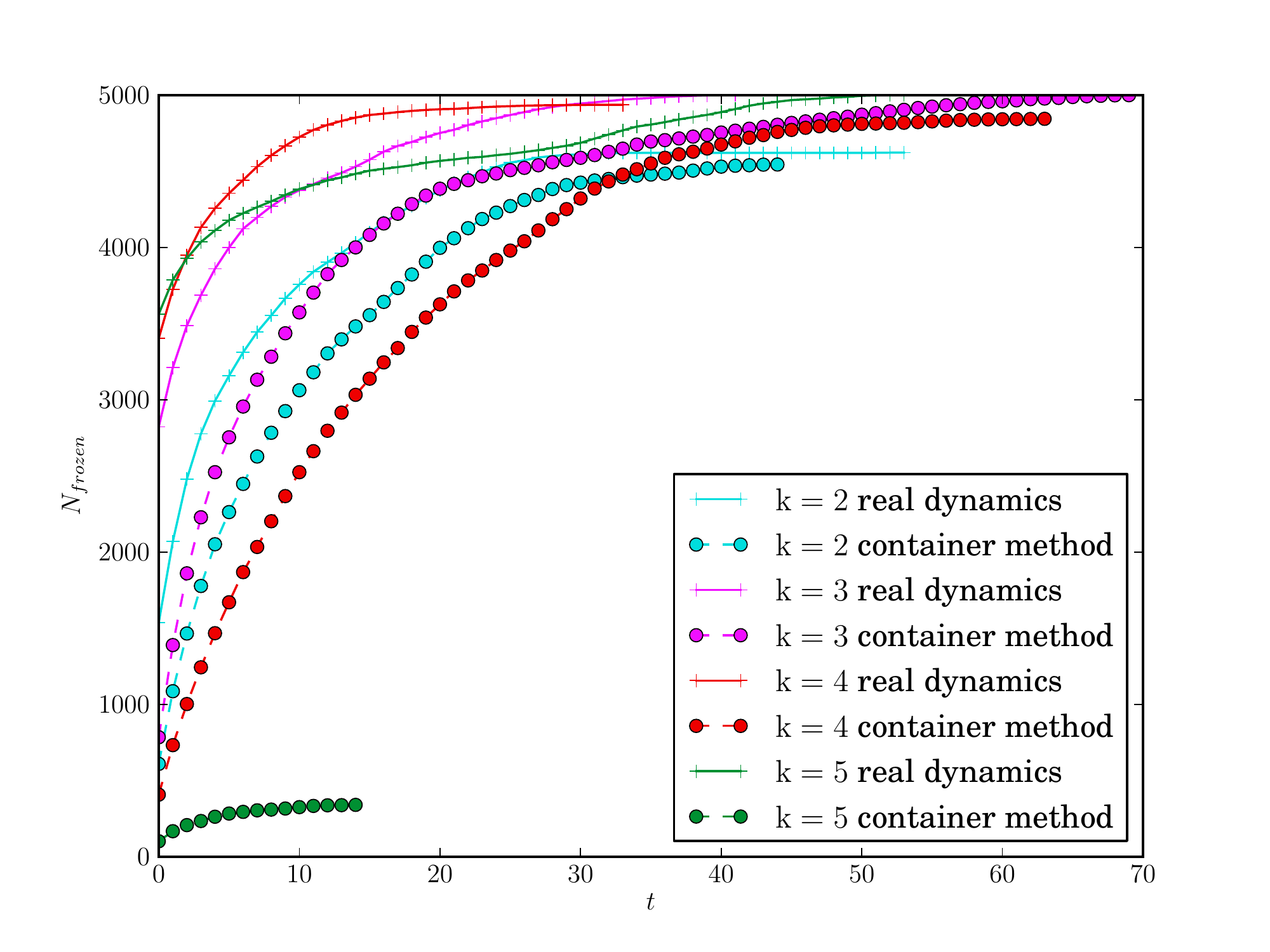}
\caption{\label{fig:ContainerCompareExample}Comparison of the freezing
  process using real dynamics and the container method with parallel
  update for four networks with $k\in\left\{2,\ldots,5\right\}$ and
  $N=5\,000$. For ``real dynamics'', a node was assumed to be frozen
  if it did not change its state any more during the remainder of the
  simulation. For the ``container method'', a node was considered
  frozen if the freezing of its input(s) caused it to freeze, with the
  initially frozen nodes being those with constant functions.}
\end{figure}

 Figure~\ref{fig:ContainerCompareExample} shows the result of such a
 comparison for $k=2,3,4,5$. The final set of frozen nodes is almost
 the same in both evaluations for $k=2,3,4$, confirming that almost
 all nodes that become frozen are part of the frozen core.  However,
 the number of nodes that are frozen at a given moment in time is
 considerably larger when all actually frozen nodes are counted and
 not only those that have become frozen because of a freezing cascade
 that begins at nodes with constant functions. The difference between
 the two simulations becomes larger for larger $k$.

In the following sections, we aim at understanding better the actual
dynamics of the formation of the frozen core. In particular, we will
investigate whether there is a qualitative difference in the freezing
dynamics and the nature of the frozen core between networks with $k<5$
and with $k \ge 5$. First, we will present computer simulations that
suggest that there is no qualitative difference. Then, we will present
an analytical calculation based on mean-field considerations that
corrobates this finding.

\section{Computer simulations of the dynamics of the formation of the frozen core}

\begin{figure}

\includegraphics[width=1\columnwidth]{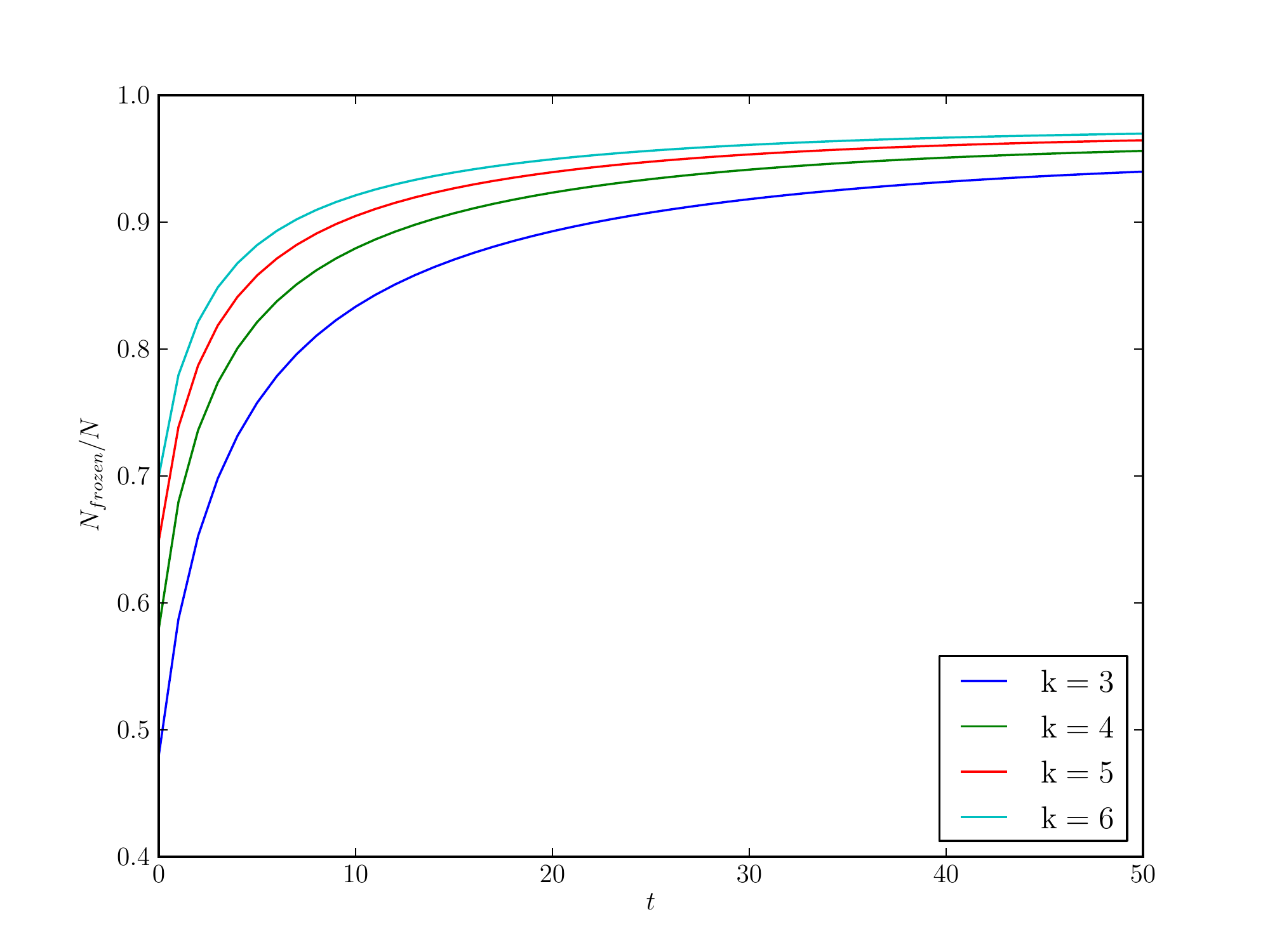}
\caption{\label{fig:containerVsReal}Proportion of frozen nodes as function of time for different values of $k$, averaged over several 1\,000 networks with $2^{14}$ nodes.}
\end{figure}

Figure~\ref{fig:containerVsReal} shows the proportion of frozen nodes
as function of time, for different values of $k$. Each curve is
averaged over several 1\,000 networks. The network size was
$N=2^{14}$, and we found that for networks as large as this the curves
do not change any more with increasing $N$. The simulations where
performed until an attractor was reached. Nodes that did not change on
the attractor were considered as frozen. If no attractor was reached
until the end of the simulation $t_{max}=\frac{N}{2}$, we considered a
node frozen after its last flip seen in the simulation. This
introduces a small finite-size effect, but as mentioned before, our
results change very little with $N$ for the values used in these
simulations. With increasing $k$, freezing becomes faster, but no qualitative
difference can be perceived between the curves for $k<5$ and $k\ge5$.

Next, we investigated whether always the same nodes become frozen.
For this purpose, we evaluated the number of nodes that become
frozen on all attractors that are reached when starting many times
from a random initial state. For $k \ge 5$, the frozen core cannot be
obtained by starting from nodes with constant functions, and for this
reason we considered it possible that the set of frozen nodes is different for
different initial conditions. We also evaluated whether a node that becomes 
frozen always freezes on the same value and whether
the propertion of nodes that freeze at the value 1 is identical to
$p$.

\begin{figure}
\includegraphics[width=1\columnwidth]{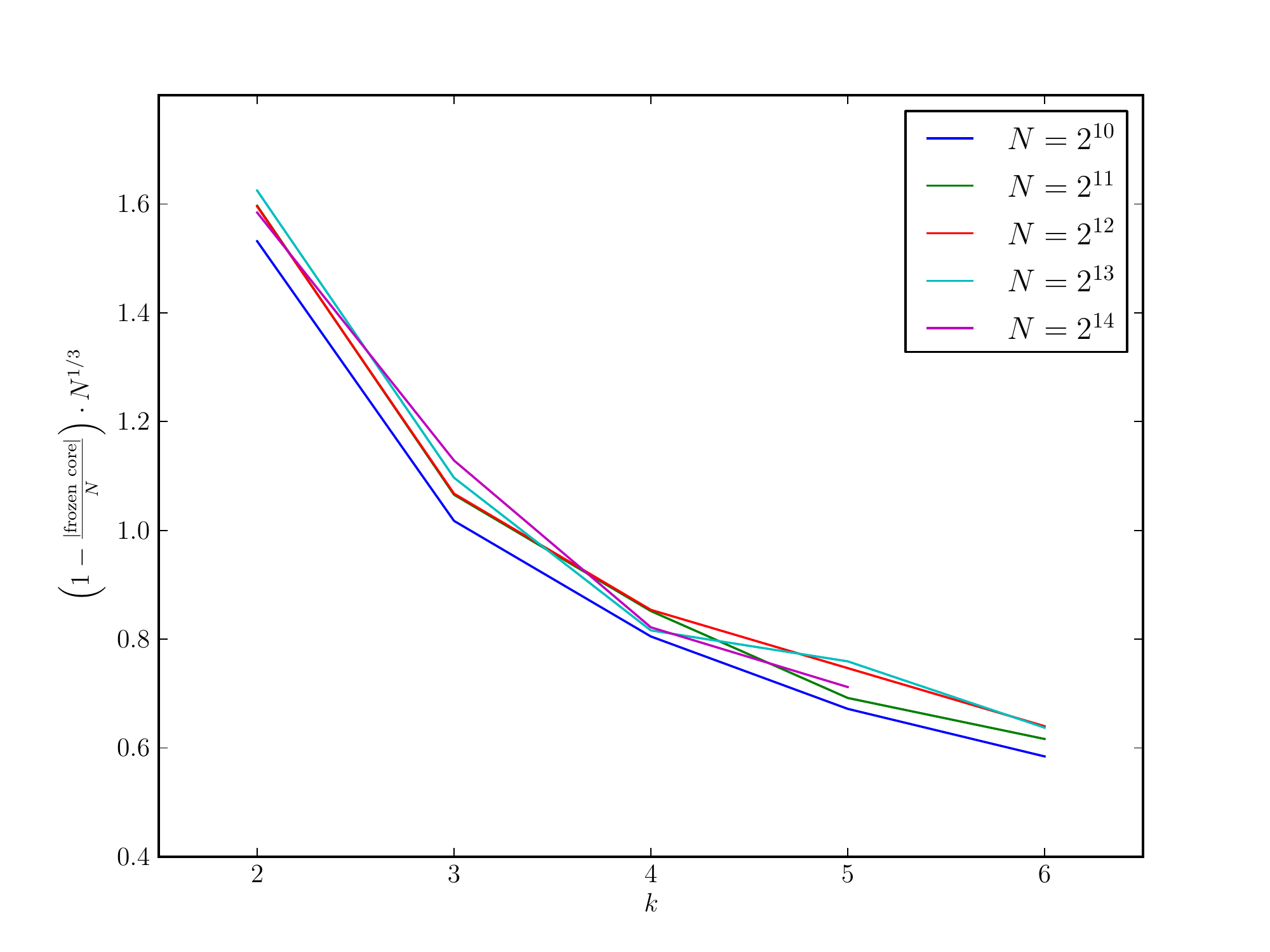}
\caption{\label{fig:robustCoreLog} Proportion of nodes that do not
  always freeze on the same value when starting from 200 different
  initial conditions, for $k=5$ and different network size $N$. The
  data are averaged over at least several hundred networks and
  are scaled with $N^{1/3}$.}
\end{figure}

Figure~\ref{fig:robustCoreLog} shows the proportion of nodes that do
not freeze for all 200 initial conditions on the same value,
averaged over at least several hundred networks, for $k=5$ and 
different values of $N$. The date are scaled with $N^{1/3}$. The
curves for different $N$ agree well with each other, indicating that
only a proportion $\sim N^{-1/3} $ of all nodes do not freeze for all
initial conditions, or do not freeze on the same value for all initial
conditions. 

\begin{figure}
 \includegraphics[width=1\columnwidth]{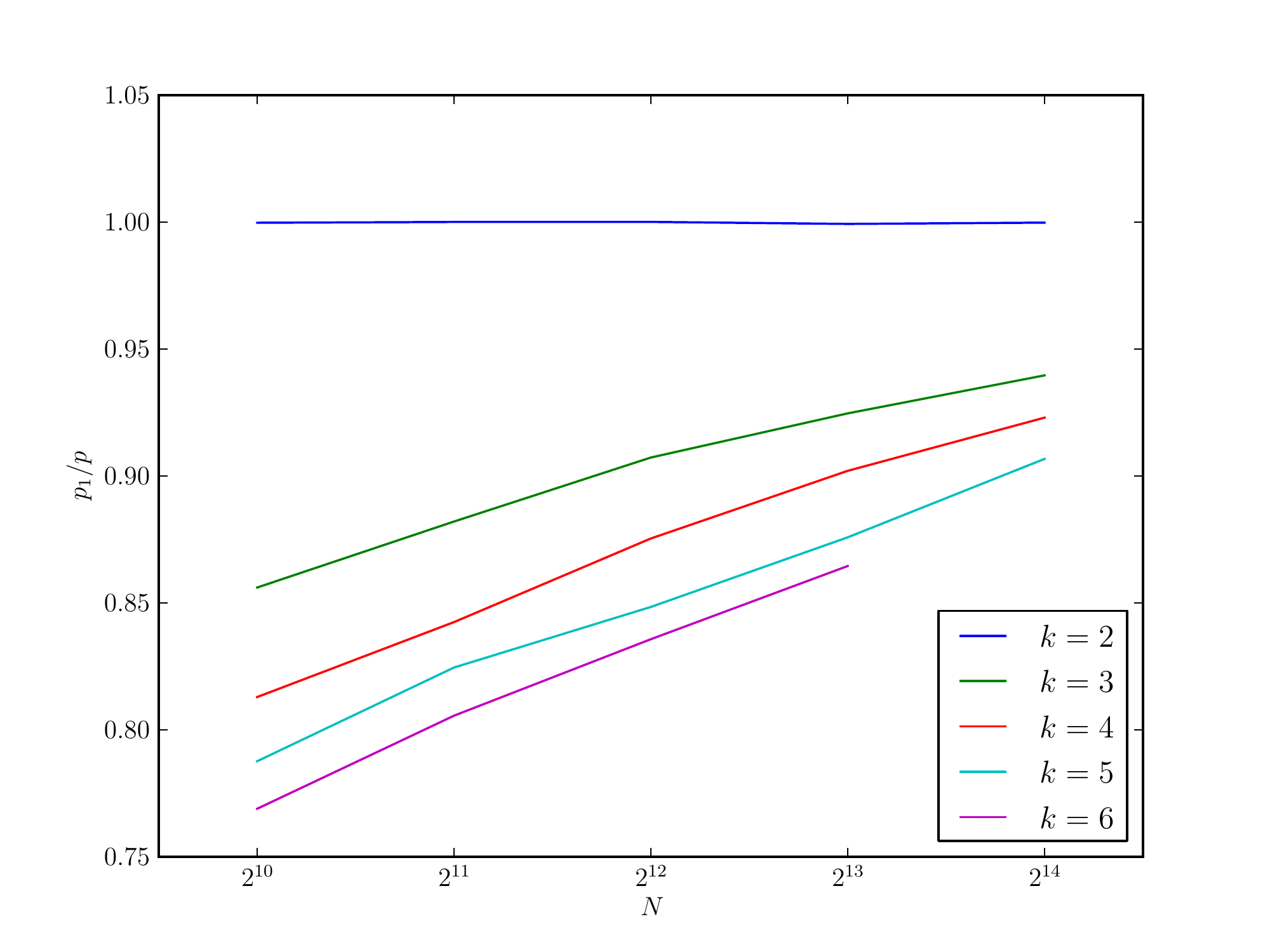}
 \caption{\label{fig:robustCorePreal} Proportion of nodes that freeze
   on the minority bit 1, divided by $p$, as a function of $N$ and
   averaged over at least several hundred networks, for different
   values of $k$.  }
\end{figure}

Figure~\ref{fig:robustCorePreal} shows the proportion of nodes that
freeze on the value 1 (which is the minority bit), divided by $p$.
This ratio approaches 1 from below with increasing $N$. The larger
$k$, the large is the deviation from 1. These data show that it is
more likely that a node freezes on its majority bit when the network
is smaller and $k$ is larger. The reason for this may be that for
larger $k$ and smaller $N$ the networks contain more short connection
loops from a node to itself. If such loops are frozen, their nodes
have to be insensitive to changes of inputs that are not part of the
loop. This means that the output of a node on the loop must be
identical for approximately $2^{k-1}$ different input states, i.e.,
for half the input states. Due to the smallness of $p$, this output
must be the majority bit.

\begin{figure}

 \includegraphics[width=1\columnwidth]{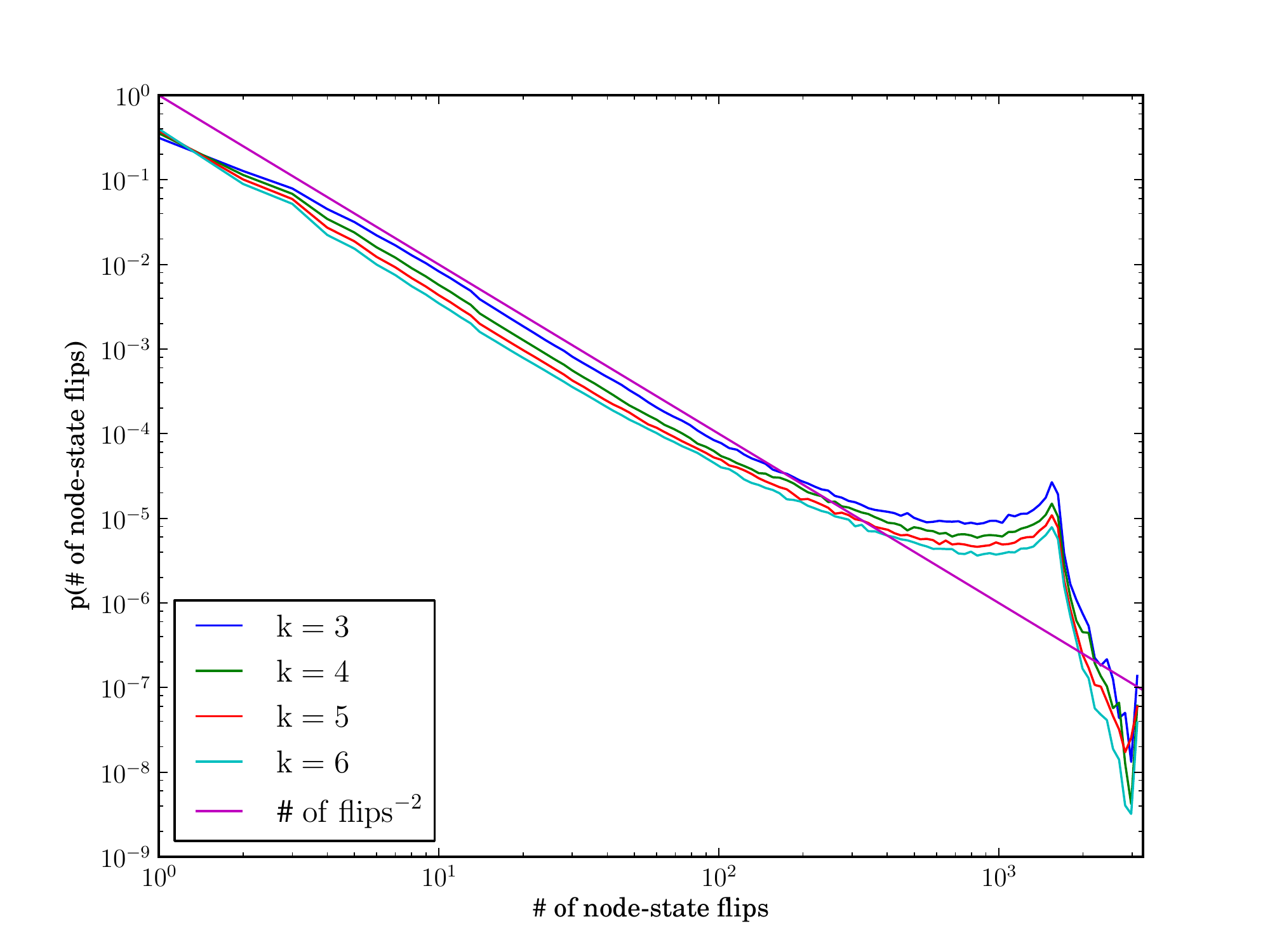}
\caption{\label{fig:plt_histNodeStateFlipDetail} 
Number of node-state flips on the transient from initial state to the attractor, averaged over at
least several thousend networks of size $N=2^{14}$. }
\end{figure}

Last, we evaluated the histogram of the number of nodes that flip at a
given moment in time when the network is not yet on an attractor. When
no attractor could be identified during the simulation time $N/2$, the
last state was assumed to be a fixed
point. Figure~\ref{fig:plt_histNodeStateFlipDetail} shows the data
obtained for $N=2^{14}$ and for different values of $k$. Again, no qualitative
change can be seen between values $k<5$ and $k \ge 5$. With increasing
$k$, the nodes freeze earlier, and the curves become lower. The peak
at the end of the curves is a finite-size effect. Its position scales
with $N$.  Before the finite-size effect sets in, the curves appear to
follow a power law with an exponent $-2$. We will confirm this
exponent in the next section, where we perform a mean-field
calculation.

\section{Mean-field  theory for the formation of the frozen core}

In the following, we perform a mean-field calculation for the
formation of the frozen core. This calculation evaluates the
probability that a nodes flips in a given time step in dependence of
the probability that at least one of its inputs has flipped in the
previous time step. This calculation neglects correlations between
nodes and between flips of the same node at different times. It is
therefore valid only when the time is short enough so that the network
has not yet reached a periodic attractor. Since the most important
relevant loop in a critical Random Boolean Network has a size of the
order of $N^{\frac{1}{3}}$ \cite{kaufmanandco:scaling}, we thus expect
finite-size effects to become visible after of the order of $N^{\frac{1}{3}}$ time
steps. 

We start from a random initial state, where each node is in the state
1 with a probability $p$. After the first time step, each node takes
the value that is prescribed by its Boolean function, given the values
of the inputs.  The probability that this is an other state as before
is $2p\left(1-p\right)=F_{0}$. (This is the probability that the node
flips from 0 to 1 or from 1 to 0.) In each of the following steps, a
node can only flip if one of its inputs has flipped in the previous
time step. We denote by $F_{t}$ the proportion of nodes that flip in
time step $t$. The probabilty that at least 1 input of a node has
flipped at time $t$, is
$\left(1-\left(1-F_{t}\right)^{k}\right)$. From this, we obtain
\begin{equation}
F_{t+1}  =  \left(1-\left(1-F_{t}\right)^{k}\right)F_{0} \, .
\label{eq:RekMeanFieldModel}
\end{equation}

The fixed points of this recursion relation are given by the condition
$F^{\star}=\left(1-\left(1-F^{\star}\right)^{k}\right)F_{0}$.  The
function on the right-hand side has its maximum slope at
$F^{\star}=0$. This slope is smaller than 1 when $k$ is smaller than
$1/(2p(1-p))$. In this case, the map (\ref{eq:RekMeanFieldModel})
converges to the fixed point 0, which means that all nodes become
frozen. In the opposite case, there is a stable fixed point at a
nonzero value of the node flip rate. At the boundary, the system is
critical, with the node-flip activity approaching zero marginally
slowly. This is equivalent to the condition formulated in
Equation~(\ref{eq:Pcrict}).

A Taylor expansion of the map
(\ref{eq:RekMeanFieldModel}) close to $F_{t}=0$ in a critical network with $F^c_0=F_0=2p(1-p)=1/k$ gives
\begin{eqnarray*}
F_{t+1} & = & \left(1-\left(1-F_{t}\right)^{k}\right)F^c_{0}\\
 & = & F_{t}-\frac{k\left(k-1\right)}{2}F_{t}^{2}F^c_{0}+\mathcal{O}\left(F_{t}^{3}\right)\, .
\end{eqnarray*}

From this, we obtain
\[
\Delta F_{t+1}=F_{t+1}-F_{t}\approx-\frac{k\left(k-1\right)}{2}F_{t}^{2}F^c_{0}
\]
Transforming this into a differential equation and integrating it leads to
\begin{equation}
F\left(t\right)  =  \frac{2}{\left(k-1\right)t}\, .\label{eq:AnalyticalModelLimit}
\end{equation}

This means that the number of flipping nodes decreases for large times
as $1/t$.  The number of nodes that flip at time $t$ for the last time
is proportional to $-\dot F(t)$, which in turn is proportional to
$1/t^2$. We have thus obtained an explanation for the exponent $-2$ found
in Figure~\ref{fig:plt_histNodeStateFlipDetail}.

\begin{figure}

 \includegraphics[width=1\columnwidth]{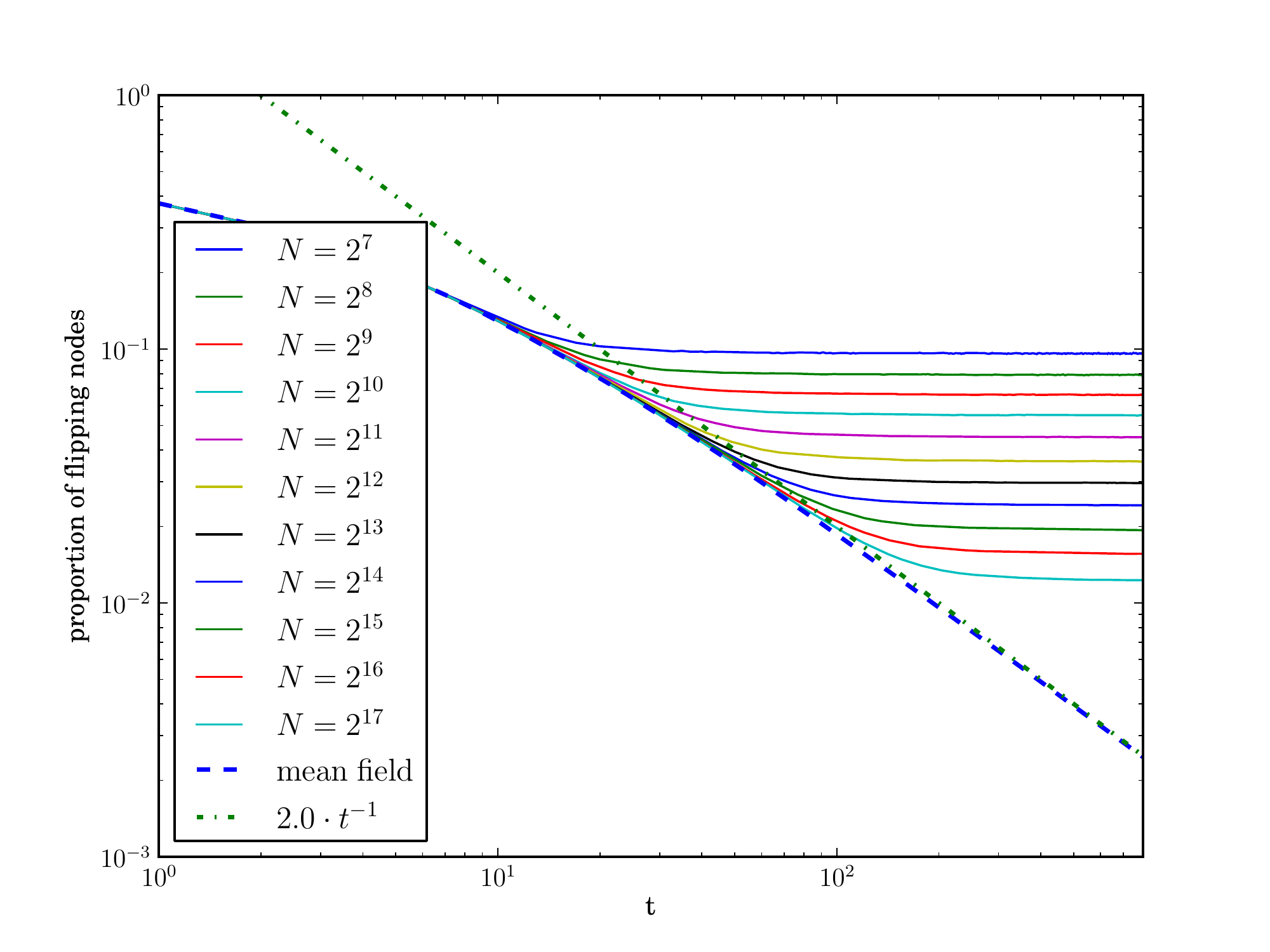}
 \caption{\label{fig:plt_flipping_k2} Comparison of mean field theory and simulation. The figure shows the proportion of flipping nodes for differerent network sizes $N$, for $k=2$, averaged over 10\,000 networks. In addition to the simulation results, this graph shows also the mean-field result and the power law (\ref{eq:AnalyticalModelLimit}) with the slope $-1$.}
\end{figure}

In order to assess the quality of our mean-field calculation, we evaluated the number of nodes that flip at each time step in real critical networks. 
Figure~\ref{fig:plt_flipping_k2} shows the results obtained for differerent network sizes $N$, for $k=2$, averaged over 10\,000
networks. In addition to the simulation results, this graph shows also the mean-field result and Equation~\ref{eq:AnalyticalModelLimit}. The graph is very similar for larger values of $k$.

One can see that up to a cutoff time the simulation data agree
perfectly with the mean-field calculation.  An evaluation of the
cutoff time reveals that it scales as $N^{\frac{1}{3}}$, which is in
agreement with our estimate of finite size effects at the beginning of
this section. After this time, the network has reached its stationary
activity level.

So far, the mean-field theory tells us only that almost all nodes
become frozen in a critical network, and that the time dependence of
the number of frozen nodes agrees with computer simulations. However,
more information can be extracted from the mean-field theory by
realizing that the recursion for $F_t$ remains identical if $F_t$ is
interpreted to be the normalized Hamming distance between two replicas
of the same network that are initiated in random initial states where
each node is 1 with probability $p$ and 0 with probability $1-p$. The
normalized Hamming distance is defined as the proportion of nodes the
states of which differ in the two replicas. The probability that the
state of a node is different in the two replicas at time $t$ is
identical to the probability that the state of at least one input was
different at time $t-1$, multiplied by the probability that a
different input leads to a different output. This leads again to the
recursion relation (12). As shown for instance in \cite{aldana2003perspectives}, the long-term behavior of the Hamming
distance can be used as a criterion for deciding whether a network is
in the frozen or chaotic phase. At the boundary, it is critical. For
critical networks, the normalized Hamming distance goes to zero, which
means that almost all nodes are frozen on the same value when the
network is started in two different random initial states. Now, if for
any two initial states almost all nodes freeze on the same value,
almost all nodes are part of the frozen core. This analytical
consideration confirms what was suggested by the simulation results
shown in Figure 5.

\section{Conclusion}

We have presented an analysis of the dynamics of the formation of
frozen nodes in critical random Boolean networks. We implemented
networks with biased functions and studied the dependence of the
properties of the frozen core on the in-degree $k$. We found that for
$k>4$ the frozen core cannot be obtained by starting from the nodes
with constant functions and by determining iteratively all nodes that
become frozen because some or all of their inputs have become
frozen. When other sets of update functions are used, this effect can
already occur at $k=3$. Nevertheless, our computer simulations of the
freezing process suggested that there is no qualitative difference
between the properties of the frozen core for $k \le 4$ and $k>4$. By
performing a mean-field calculation for the number of nodes that flip
as a function of time, we could calculate two power laws that are
observed in the computer simulations, and we could show that the
dynamics of the formation of the frozen core can be captured correctly
by neglecting correlations between nodes and between subsequent flips
of the same node. 

Furthermore, our computer simulations showed that irrespective of the
initial state always the same nodes freeze, apart from a fraction
proportional to $N^{-1/3}$, and that these nodes always freeze in the
same state. For $k \le 4$, this result is obtained by starting from
nodes with constant functions. For $k>4$, this result points to the
existence of large groups of nodes that remain frozen once they are
fixed on specific values. We confirmed this observation with a
mean-field calculation. 

Due to the universality of critical behavior, we expect our results to
hold for random Boolean networks with other sets of Boolean functions
apart from biased functions, and for networks where not all nodes have
the same in-degree $k$, as long as the second moment of the
distribution of $k$ values is finite \cite{lee2008broad}.

To conclude, viewing the freezing process of critical random Boolean
networks as an avalanche that starts at nodes with constant functions,
is an unnecessary limitation that does not capture the universality of
the freezing process. Rather, as revealed by mean-field theory,
freezing is ultimately due to the insufficient sensitivity of the
nodes to changes in their inputs. A critical network with biased
functions and a large value of $k$ contains few constant functions,
but its value of $p$ is so small that the number of node-state flips
decreases fast, and so does the difference between two replicas of the
same network.

\section*{acknowledgements}

This work was supported by the DFG under grant number Dr300/4-2

\section*{References}{}

\bibliographystyle{unsrt}
\bibliography{literatur}

\end{document}